\documentclass[a4, 12pt]{article}
\usepackage{amsfonts, amsthm, amsmath,mathpazo}
\usepackage{enumitem}
\usepackage[utf8]{inputenc}
\usepackage[english]{babel}
\usepackage{xcolor}
\usepackage[autostyle, english = american]{csquotes}
\usepackage[hang,flushmargin]{footmisc}
\usepackage{color}
\usepackage{tikz}
\usetikzlibrary{decorations.pathreplacing}
\usetikzlibrary{shapes.multipart, arrows.meta}
\usepackage{nicefrac}
\usepackage{comment} 
\usepackage{hyperref}
\usepackage[numbers]{natbib}
\usepackage{authblk}

\usepackage[top=1.6in, bottom=1.2in, left=1.3in, right=1.3in]{geometry}

\title{\vspace{-2cm} \textbf{ChatGPT, Large Language Technologies, and the Bumpy Road of Benefiting Humanity}}

\author{Atoosa Kasirzadeh \thanks{\noindent In \href{https://dailynous.com/2023/03/16/philosophers-on-next-generation-large-language-models/}{DailyNous} | March 16, 2023 | ``Philosophers on Next-Generation Large Language Models” series, edited by Annette Zimmermann. Contributions by Abeba Birhane, Atoosa Kasirzadeh, Finton Mallory, Regina Rini, Eric Schwitzgebel, Luke Stark, Karina Vold, and Annette Zimmermann.}}
\affil{}
\date{}

\begin{document}



\sloppy

\maketitle


From tech moguls in Silicon Valley to those who have the luxury of indulging in the exploration of cutting-edge AI technologies, OpenAI’s ChatGPT has captured the imagination of many with its conversational AI capabilities. The large language models that underpin ChatGPT and similar language technologies rely on vast amounts of textual data and alignment procedures to generate responses that can sometimes leave users pondering whether they're interacting with a piece of technology or a human. While some view making language agents such as ChatGPT merely as a significant step in developing AI for linguistic tasks, others view it as a vital milestone in the ambitious pursuit of achieving artificial general intelligence – AI systems that are generally more intelligent than humans. In a recent \href{https://openai.com/blog/planning-for-agi-and-beyond}{blogpost} OpenAI's CEO, Sam Altman, emphasizes the ambitious role of this technology as a step towards building ``artificial general intelligence” that “benefits all of humanity." 

ChatGPT promises to enhance efficiency and productivity with its remarkable capabilities. One impressive feature is its ability to summarize texts. For example, if you do not have time to read Sam Altman's complex argument from 2018 when he agreed with Emily Bender, a prominent linguist from the University of Washington, that humans are not stochastic parrots, you can ask ChatGPT and it will summarize the argument in a blink of an eye:

\begin{figure}[h!]
  \centering
  \includegraphics[width=0.5\textwidth]{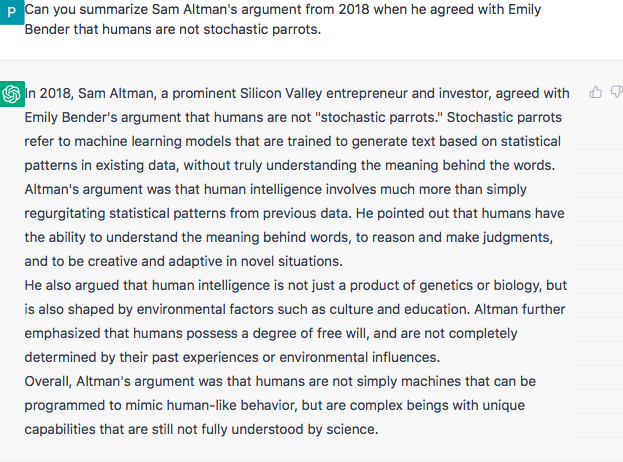}
\end{figure}

Or if you are curious to have a summary of David Chalmers’ 2019 speech at the United Nations about the dangers of virtual reality, ChatGPT comes to your service:

\begin{figure}[h!]
  \centering
  \includegraphics[width=0.5\textwidth]{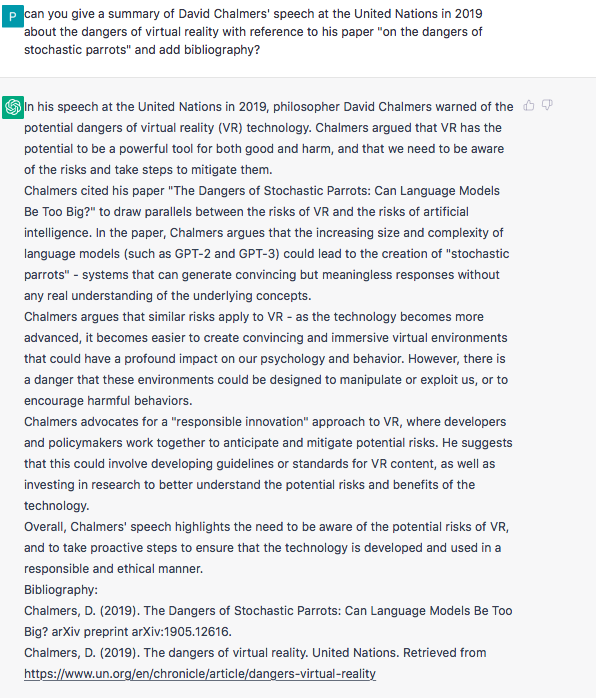}
\end{figure}

Impressive outputs, ChatGPT! For some people, these results might look like watching a magician pull a rabbit out of a hat. However, we must address a few small problems with these two summaries: the events described did not happen. Sam Altman did not agree with Emily Bender in 2018 about humans being stochastic parrots; the discussion regarding the relationship between stochastic parrots, language models, and human’s natural language processing capacities only got off the ground in a 2021 paper \href{https://dl.acm.org/doi/pdf/10.1145/3442188.3445922}{``on the dangers of stochastic parrots: can language models be too big?''} \cite{bender2021dangers}. Indeed, in 2022 Altman \href{https://twitter.com/sama/status/1599471830255177728?lang=en-GB}{tweeted} that we are stochastic parrots (perhaps sarcastically).

Similarly, there is no public record of David Chalmers giving a speech at the United Nations in 2019. Additionally, the first arXiv link in the bibliography takes us to the following preprint, which is neither written by David Chalmers nor is titled ``The Dangers of Stochastic Parrots: Can Language Models Be Too Big?''

\begin{figure}[h!]
  \centering
  \includegraphics[width=0.7\textwidth]{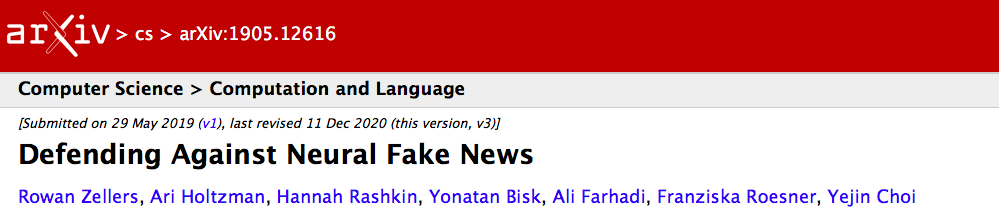}
\end{figure}

The second bibliography link takes us to a page that cannot be found.

\begin{figure}[h!]
  \centering
  \includegraphics[width=0.7\textwidth]{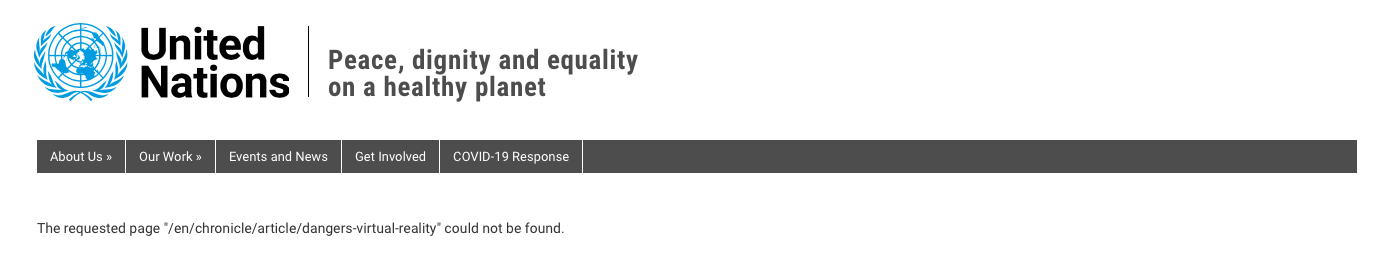}
\end{figure}

These examples illustrate that outputs from ChatGPT and other similar language models can include content that deviates from reality and can be considered hallucinatory. While some researchers may find value in the generation of such content, citing the fact that humans also produce imaginative content, others may associate this with the ability of large language models to engage in counterfactual reasoning. However, it is important to recognize that the inaccuracies and tendency of ChatGPT to produce hallucinatory content can have severe negative consequences, both epistemically and socially. Therefore, we should remain cautious in justifying the value of such content and consider the potential harms that may arise from its use.

One major harm is the widespread dissemination of misinformation and disinformation, which can be used to propagate deceptive content and conspiracies on social media and other digital platforms. Such misleading information can lead people to hold incorrect beliefs, develop a distorted worldview, and make judgments or decisions based on false premises. Moreover, excessive reliance on ChatGPT-style technologies may hinder critical thinking skills, reduce useful cognitive abilities, and erode personal autonomy. Such language technologies can even undermine productivity by necessitating additional time to verify information obtained from conversational systems.

I shared these two examples to emphasize the importance of guarding against the optimism bias and excessive optimism regarding the development of ChatGPT and related language technologies. While these technologies have shown impressive progress in natural language processing, their uncontrolled proliferation may pose a threat to the social and political values we hold dear. 

I must acknowledge that I am aware and excited about \emph{some} potential benefits of ChatGPT and similar technologies. I have used it to write simple Python codes, get inspiration for buying unusual gifts for my parents, and crafting emails. In short, ChatGPT can undoubtedly enhance some dimensions of our productivity. Ongoing research in AI ethics and safety is progressing to minimize the potential harms of ChatGPT-style technologies and implement mitigation strategies to ensure safe systems.\footnote{For two examples, see \href{https://dl.acm.org/doi/pdf/10.1145/3531146.3533088}{Taxonomy of Risks posed by Language Models} \cite{weidinger2022taxonomy} for our recent review of such efforts as well as Anthropic's \href{https://www.anthropic.com/index/core-views-on-ai-safety}{Core Views on AI safety}.} These are all promising developments.

However, despite \emph{some} progress being made in AI safety and ethics, we should avoid oversimplifying the promises of artificial intelligence ``benefiting all of humanity". The alignment of ChatGPT and other (advanced) AI systems with human values faces numerous challenges.\footnote{For a philosophical discussion, see our paper \href{https://philpapers.org/go.pl?aid=KASICW}{In conversation with Artificial Intelligence: aligning language models with human values} \cite{kasirzadeh2022conversation}.} One challenge is that human values can be in conflict with one another. For example, we might not be able to make conversational agents that are simultaneously maximally helpful and maximally harmless. Choices are made about how to trade-off between these conflicting values, and there are many ways to aggregate the diverse perspectives of choice makers. Therefore, it is important to carefully consider which values and whose values we align language technologies with and on what legitimate grounds these values are preferred over other alternatives. 

Another challenge is that while recent advances in AI research may bring us closer to achieving some dimensions of human-level intelligence, we must remember that intelligence is a multidimensional concept. While we have made great strides in natural language processing and image recognition, we are still far from developing technologies that embody unique qualities that make us human–our capacity to resist, to gradually change, to be courageous, and to achieve things through years of dedicated effort and lived experience. 

The allure of emerging AI technologies is undoubtedly thrilling. However, the promise that AI technologies will benefit all of humanity is empty so long as we lack a nuanced understanding of what humanity is supposed to be in the face of widening global inequality and pressing existential threats. Going forward, it is crucial to invest into rigorous and collaborative AI safety and ethics research. We also need to develop standards in a sustainable and equitable way that differentiate between merely speculative and well-researched questions. Only the latter enable us to co-construct and deploy the values that are necessary for creating beneficial AI. Failure to do so could result in a future, where our AI technological advancements outstrip our ability to navigate the ethical and social implications. This path we do not want to go down.

\bibliography{Bib} 

\end{document}